\documentclass[aps, prl, superscriptaddress, reprint]{revtex4-1}

\usepackage{graphicx}

\begin{document}

\title{Anomalous polarization-dependent transport in nanoscale double-barrier 
superconductor/ferromagnet/superconductor junctions }

\author{Madalina \surname{Colci}}
\email{colci@illinoisalumni.org}
\affiliation{Department of Physics and Frederick Seitz Materials Research Laboratory, University of Illinois at Urbana-Champaign, Urbana, Illinois 61801, USA}
\author{Kuei \surname{Sun}}
\affiliation{Department of Physics and Frederick Seitz Materials Research Laboratory, University of Illinois at Urbana-Champaign, Urbana, Illinois 61801, USA}
\affiliation{Department of Physics, University of Cincinnati, Cincinnati, Ohio 45221, USA}
\author{Nayana \surname{Shah}}
\affiliation{Department of Physics, University of Cincinnati, Cincinnati, Ohio 45221, USA}
\author{Smitha \surname {Vishveshwara}}
\author{Dale J. \surname{Van Harlingen}}
\affiliation{Department of Physics and Frederick Seitz Materials Research Laboratory, University of Illinois at Urbana-Champaign, Urbana, Illinois 61801, USA}

\date{\today}

\begin{abstract}
We study the transport properties of nanoscale superconducting (S) devices in which two superconducting electrodes are bridged by two parallel ferromagnetic (F) wires, forming an SFFS junction with a separation between the two wires less than the superconducting coherence length. This allows crossed Andreev reflection to take place. We find that the resistance as a function of temperature exhibits behavior reminiscent of the re-entrant effect and, at low temperatures and excitation energies below the superconducting gap, the resistance corresponding to antiparallel alignment of the magnetization of the ferromagnetic wires is higher than that of parallel alignment, in contrast to the behavior expected from crossed Andreev reflection. We present a model based on spin-dependent interface scattering that explains this surprising result and demonstrates the sensitivity of the junction transport properties to interfacial parameters.
 \end{abstract}

\maketitle

An emerging focus of mesoscale quantum physics is the exploration of interfaces between materials with distinct electronically correlated phases. An interesting example is hybrid heterostructures made of superconductors (S) and ferromagnets (F) which have revealed unexpected phenomena arising from the interplay of Cooper pairing in the superconductor and the pair-breaking effects of the exchange field in the ferromagnet \citep{Ryazanov01, Kontos02, Blum02, Giroud, Lawrence99, Petrashov99, Keizer, Birge}. Advances at the nanoscale have opened the possibility to study non-local effects emerging in a multiterminal system made of a superconductor coupled to two ferromagnetic or two normal metal (N) wires separated by a distance less than the superconducting coherence length $\xi_S$ \citep{Beckmann04, Russo05, Kleine}.  In this configuration, non-local correlations between electrons with opposite spins are predicted to occur as a result of crossed Andreev reflection (CAR)  \citep{Byers, Deutscher, Recher}, a process that can be described as the splitting of Cooper pairs into constituent entangled electrons in separate leads. Conventional spin-singlet superconductors are a natural source of entangled electrons, making such structures highly attractive from the perspective of quantum information \citep{Burkard}; ultimately the read-out of these entangled states may require spin-selective measurements that can be achieved by coupling to ferromagnetic leads.

\begin{figure}
\includegraphics[width=0.8\linewidth]{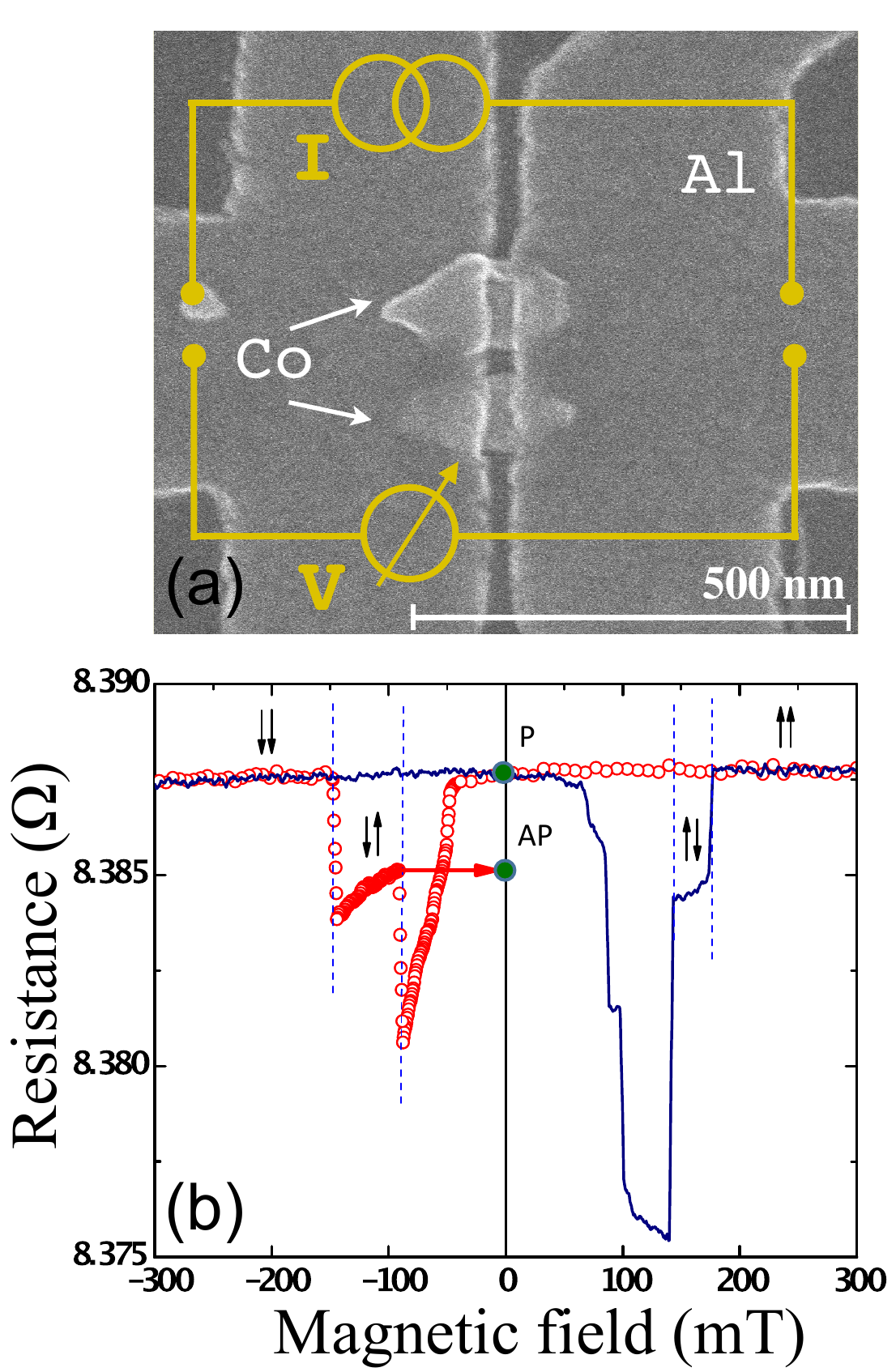}
\caption{(Color online) (a) Scanning electron micrograph of a double junction (device A) in which two aluminum (Al) electrodes are bridged by two closely spaced cobalt (Co) wires. The schematic of current injection and voltage detection is also shown. (b) Magnetoresistance curve for device B as the magnetic field is swept from negative to positive values (solid line) and from positive to negative values (circles). The black arrows label the four magnetization configurations of the two wires.}
\label{sffs1}
\end{figure}

In this Rapid Communication we investigate the interplay between non-local pair correlations in superconductors and the effect of ferromagnetic interfaces by studying transport measurements in an SFFS geometry in which the S electrodes are bridged by two F wires separated by a distance $L$ much smaller than $\xi_S$  [see Fig. \ref{sffs1}(a)]. We find that as the temperature is lowered below the critical temperature  $T_c$ of the superconductor, the resistance of the antiparallel (AP) alignment of the magnetization of ferromagnetic wires becomes larger than that of the parallel (P) case, a behavior in surprising contrast to that expected from crossed Andreev reflection. To analyze this result we develop a theoretical model that incorporates a spin-dependent scattering potential at the two FS interfaces. Our study considers spin-active interfaces in SFF modeling, which prove critical for explaining the experimental data and also lead to a wide range of behavior depending on the parameters of the system.

In the following we show measurements on two devices, labeled A and B. Figure \ref{sffs1}(a) shows a scanning electron microscopy image of one of our double junctions. All samples are fabricated in two steps by electron-beam lithography and thermal evaporation. First, we define two ferromagnetic wires by evaporating 20 nm of cobalt (Co) onto a patterned oxidized silicon substrate. We determine the low-temperature material parameters from control structures fabricated on the same chip as the SFFS devices. For device A, the  Co resistivity and the corresponding diffusion coefficient are 70 $\mu \Omega$ $ $cm and 0.7 cm$^2$/s. For device B, with higher purity Co, the same parameters are 16 $\mu \Omega$ $ $cm and 3.1 cm$^2$/s. The wires are 200 nm long and have different widths (90 nm and 120 nm) to ensure different coercive fields. The edge-to-edge distance between the Co wires is 30$-$40 nm, substantially smaller than the dirty limit $\xi_S$ of aluminum (Al), which is 140 nm in our samples. We next define the two superconducting electrodes by electron-beam lithography, followed by ion-mill cleaning of the Co surface and deposition of 60 nm of Al with a resistivity of 2.3 $\mu \Omega$ $ $cm and an elastic mean free path of 16.5 nm. The contacts between the Al electrodes and the Co wires have low resistance and an overlap smaller than $\xi_S$ that prevents the suppression of the superconductivity in the contact region by the exchange field of the ferromagnet. The gap between the electrodes is approximately 40 nm, a distance smaller than $\xi_S$ but much larger than the estimated penetration length of superconducting correlations in the ferromagnet, which is 0.6 nm and 1.3 nm in the two devices.

We measure the electron transport properties in a dilution refrigerator with a base temperature of 8 mK. The circuit shown in Fig.~\ref{sffs1}(a) makes a four-terminal resistance measurement using a low-frequency ac resistance bridge with an rms excitation current  $I_{\rm ac}$. We first characterize the samples by magnetoresistance measurements in the normal state of the Al electrodes at $T$ = 4 K by sweeping an in-plane magnetic field $B$ along the axis of the Co wires, and measuring the resistance with  $I_{\rm ac}$ = 10 $\mu$A. Figure~\ref{sffs1}(b) shows characteristic changes in resistance for our devices when the magnetization of the Co wires first rotates and then abruptly changes direction as a function of the applied field. This is typical  anisotropic magnetoresistance (AMR) behavior in which the resistance is dependent upon the angle between the direction of current flow and the orientation of the magnetization \citep{McGuire}.

For each sweep of the magnetic field the sharp changes in resistance correspond to the reversal of the magnetization of each of the two wires when they reach their coercive field values. The devices' magnetic configuration switches first to the antiparallel alignment and, as the field is increased further, to the parallel alignment. The alignment will remain unchanged as the field is decreased to zero; this is how the P state is set before we cool the device below the superconducting transition temperature. In order to prepare the AP state, we reverse the field sweep direction when only one of the wires has switched; the AP alignment remains stable as the field is decreased to zero.

 We note that we can distinguish an AP state with a lower resistance value than the P state even though the AMR resistance should be the same when the direction of magnetization relative to the current  is 0 and $\pi$. This is an artifact of the magnetic field sweep rate combined with the similar switching fields of the two wires: The AP state cannot reach the same resistance value as the P state because the wires reverse almost simultaneously. This is best seen on the negative branch of the field sweep. The positive branch displays a broad drop in resistance with several steps indicative of a multiple domain structure in the wider Co strip. In this case the magnetoresistance has contributions from domain wall resistance, which is why it drops so far before settling into the AP state.

\begin{figure}
\includegraphics[width=1.0\linewidth]{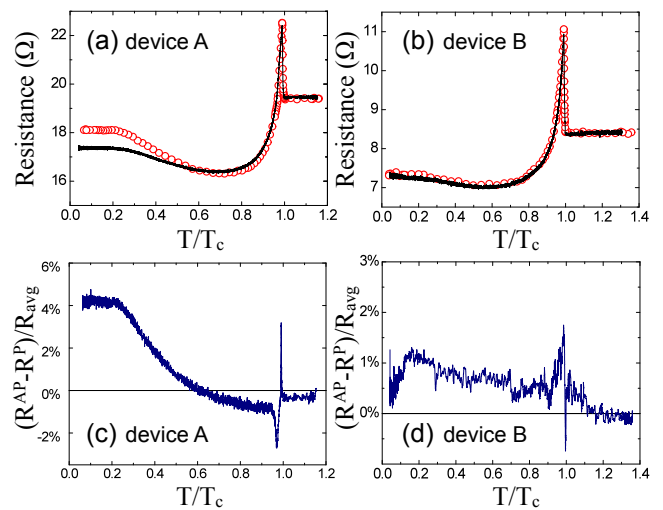}
\caption{(Color online) (a), (b) Temperature dependence of the resistance of our SFFS devices as they are cooled below the superconducting transition temperature $T_c$. Solid line: Parallel magnetization alignment. Symbols: Antiparallel alignment. (c), (d) Difference in resistance between AP and P magnetization alignment for devices A and B as they are cooled through the superconducting transition.}
\vspace{0.0in}
\label{SFFS-Temp} 
\end{figure}

 We cool the devices below $T_c$ for P and AP magnetization alignments, recording the resistance as the temperature is lowered [Figs. \ref{SFFS-Temp}(a) and \ref{SFFS-Temp}(b)]. The signal is measured using a small excitation $I_{\rm ac} = 100$ nA to avoid quasiparticle injection above the superconducting gap at the lowest temperatures. In both magnetic states we observe a sharp increase in resistance below $T_c$ due to charge  \citep{clarke} and spin imbalance \citep{johnson}, followed by a decrease to a minimum value at approximately $0.65 T_c$. As the temperature is lowered further, the resistance starts increasing again and, at $0.25 T_c$, it saturates to a value close to the normal state resistance. The resistance never drops to zero, and a supercurrent is not resolved. At the lowest temperature we measure differential resistance curves (not shown here) as a function of bias current for the two magnetization states. We observe a peak at zero bias, and a symmetric structure of peaks at higher bias currents, which we attribute to multiple Andreev reflections (MARs)  \citep{mars}. The presence of MARs in our devices indicates that transport across our junctions is phase coherent. We defer a detailed study of these features to a later paper.
	
The most important result of our experiment is that at the lowest temperature the AP state has a noticeably higher resistance than the P state. Intuitively, this result is surprising: The antiparallel configuration should have the lowest resistance since Cooper pairs encounter opposite exchange fields at the SF interface, which should have a smaller pair-breaking effect than in the P-aligned configuration.

In Figs.~\ref{SFFS-Temp}(c) and \ref{SFFS-Temp}(d), we plot the relative resistance $\delta R$ as the difference in resistance of the two states $R^{\rm AP} - R^{\rm P}$ normalized by their average $R_{\rm avg} = (R^{\rm AP} + R^{\rm P})/2$. This function has a non-monotonic temperature dependence, and the behavior is similar but not identical in the two devices. In device A the AP resistance increases as the temperature is lowered until the saturation point is reached at $0.23 T_c$; the P-AP crossover temperature is $0.6 T_c$. By contrast, in device B we see the crossover occurs at $T_c$; as the temperature is lowered, the behavior is similar to that of device A, but at the same temperature where the device A curve flattens out, the P-state resistance of device B starts to increase, dramatically reducing the difference in resistance between the two magnetic states at the lowest temperature.

Our devices have similar dimensions, with small variations that cannot explain the large difference in $\delta R$ between device A and B. However, device B is likely to have a higher polarization due to the use of higher-purity cobalt. We note that the low-temperature split between P and AP states is decreased in this device.

There are three other experimental studies \citep{Luo09, Lin11, Almog11} of double junctions with the same geometry as ours, but with larger separation between the S electrodes and between the F wires. The measurements in Ref. \citep{Luo09, Lin11} show $\delta R<$ 0, a behavior explained by spin accumulation in the S electrodes. The effect in Ref. \citep{Luo09} decreases as the temperature is lowered, in marked contrast to our observations.

Results similar to ours were recently reported by Almog {\it et al.} \cite{Almog11}. Unlike in our junctions, their SF interface is not homogeneous due to pinholes in the native oxide layer on top of the Co wires. Further complicating the picture is interfacial domain formation during magnetization reversal, known to occur in exchange-biased CoO/Co bilayers \cite{Welp03}. Therefore, direct comparison with our devices may not be appropriate. However, the presumed small junction area and the low polarization of the wires may be the main factors for the similar behavior observed. Their theoretical explanation cannot account for our results: Spin-triplet superconductivity would penetrate the F layers over a long length \cite{Bergeret}, which in our wires is greater than the separation between the S electrodes; this would allow supercurrent to flow in the P state.

 \begin{figure}
\includegraphics[width=1.0\linewidth]{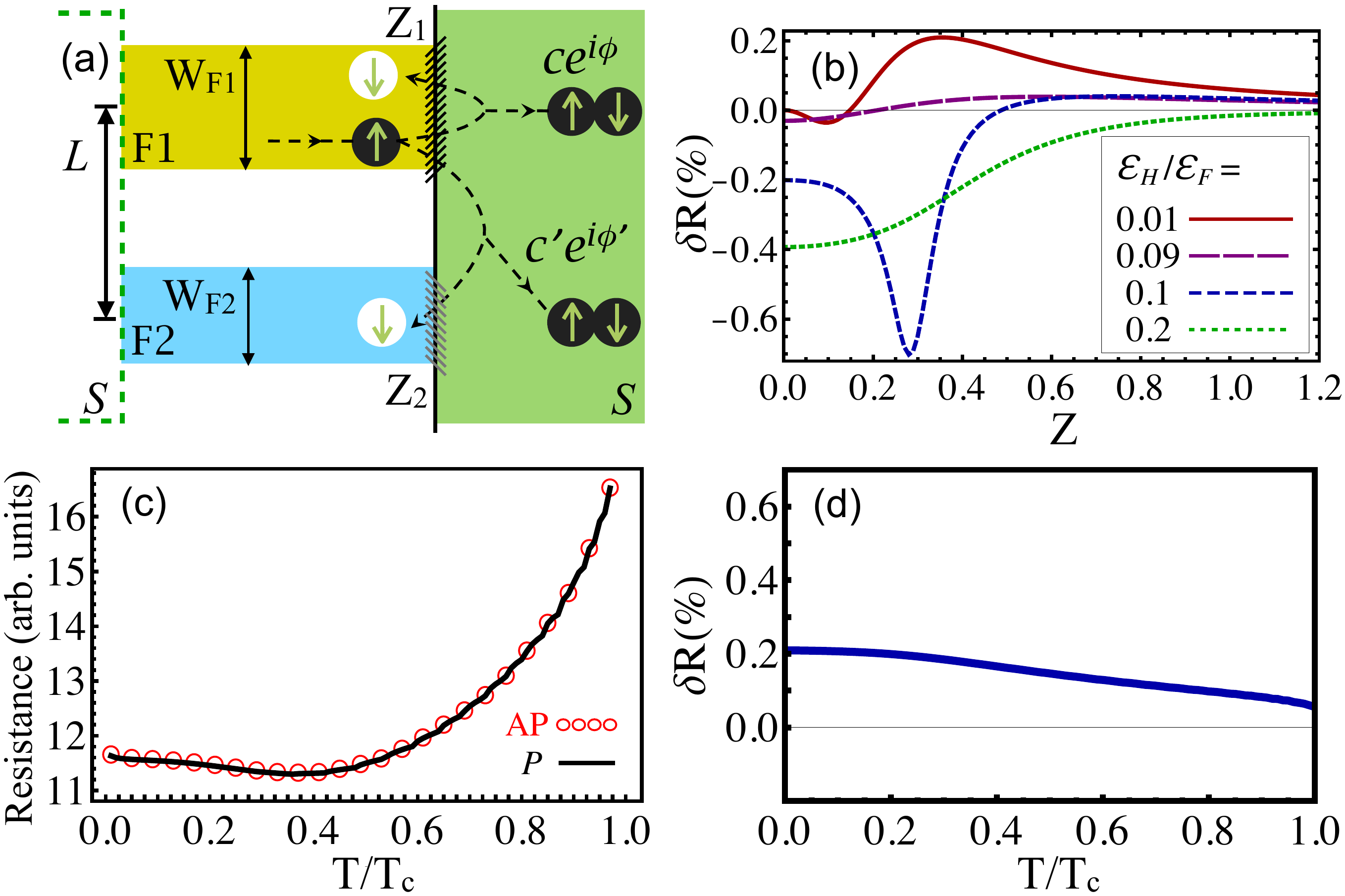}
\caption{(Color online) (a) Cartoon representation of the spin-dependent processes at one SFF interface. The structure has a superconducting electrode S in contact with two ferromagnetic wires F1 and F2, with widths $W_{\rm F1}$ and $W_{\rm F2}$ and separation $L$ between them. The dashed arrows indicate the AR and CAR processes of an incident spin-up electron (black disk in F1) which result in the reflection of spin-down holes (white disks in F1 and F2) and transmission of Cooper pairs (black disk pairs in S) with different amplitudes $c e^{i\phi}$ and $c' e^{i\phi'}$ due to the spin-dependent interface parameters $Z_1$ and $Z_2$. (b) Relative resistance $\delta R$ as a function of $Z$ for different  polarizations $\varepsilon_{H}/\varepsilon_{F}$ at $T=0$. (c) Temperature dependence of the resistance for the parallel (black solid curve) and antiparallel (red circles) configuration for $|Z|=0.35$ and $\varepsilon_H/\varepsilon_F=0.01$. (d) Temperature dependence of  $\delta R$ for the same parameters as in (c).} 
\label{theory} 
\end{figure} 

To understand the relative resistance between P and AP orientations, we model the essential features of one interface of our SFFS devices using a modified Blonder-Tinkham-Klapwijk (BTK) treatment  for such structures made of a spin-singlet superconductor in contact with two ferromagnetic wires \citep{Yamashita03}, as depicted in Fig.~\ref{theory}(a). The ferromagnets F1 and F2 are modeled as partially polarized Fermi liquids having a Fermi energy $\varepsilon_F$ and a Zeeman splitting $\varepsilon_H$ of the two spin components along a fixed direction, thus distinguishing a major (M) and minor (m) spin species in each wire. 

For an electron injected from the F side, the scattering processes at the interface are quasiparticle transmission in S, normal reflection, Andreev reflection (AR), crossed Andreev reflection (CAR), and elastic cotunneling (EC). For energies below the superconducting gap, transport is possible by means of all but the first process. In the AR process \citep{Andreev} [Fig.~\ref{theory}(a)] a spin-up electron incident on the FS interface is retroreflected as a spin-down hole, creating a Cooper pair in the superconductor. For CAR, the incoming electron is reflected as a hole in the other F wire. For EC, another nonlocal process, the electron in F1 is transmitted as an electron in F2 without change of spin. The BTK treatment accounts for the AR and CAR  processes as a linear superposition [Fig.~\ref{theory}(a)] and the total current is altered by their interference. 

 Intuitively, CAR is enhanced in the AP configuration because the majority species in the two wires have the opposite spin necessary to form a Cooper pair in S \citep{Deutscher}; therefore, the AP state should have a lower resistance. Indeed, previous theoretical treatments have shown lower resistance for the AP case for all values of model parameters \citep{Yamashita03}.
 
 In our model \citep{KSun} we introduce another element that changes this behavior: spin-dependent scattering at each FS interface characterized by different potential barrier strengths $Z_{\rm M}$ and $Z_{\rm m}$ for the majority and minority spin species. Such spin-dependent scattering arises naturally from the magnetic properties of the barrier, such as the Zeeman splitting and spin-orbital coupling. Indeed, it has been shown to play a major role for a single interface \citep{deJong95, Kastening09, Kupferschmidt10}. Thus, in contrast to the results of the spin-independent interface model \citep{Yamashita03}, we find that in the AP alignment, where each spin species is a majority carrier in one wire and a minority carrier in the other wire, $Z_{\rm M}$ and $Z_{\rm m}$ cause the two spin species to scatter differently through each FS interface. Within this setting we solve the Bogoliubov-de Gennes equation to obtain the current-voltage relationship as a function of the barrier strengths. Crucially, for a specific set of parameters, interference between the scattering processes enhances the resistance in the AP alignment compared to that of the P alignment, concurring with our experimental results.
 
In Figs.~\ref{theory}(b)$-$\ref{theory}(d) we present the results of our calculation for parameters that are optimal for comparison to our experiment. We focus on the case of barrier strength $Z_{\rm M}$ = $-Z_{\rm m}$ = $Z$, a choice that reflects the effect of the Zeeman potential at SF interfaces \citep{KSun, Kastening09,Kupferschmidt10}. The minus sign leads to the majority and minority spins acquiring opposite phase shifts upon interface scattering \citep{Tokuyasu}. In this specific case we observe a significant parameter region where $\delta R > $ 0, as in our experiment, which is completely absent for the spin-independent case.

In Fig.~\ref{theory}(b) we plot $\delta R$ as a function of $Z$ for different polarizations $\varepsilon_H/\varepsilon_F$. The sign of  $\delta R$  is highly sensitive to the values of model parameters.  Notably, for low polarization similar to our experiment, the interference arising from the spin-dependent phase shift dominates and we find that the AP configuration can have a higher resistance than the P case over a wide range of $Z$ values. In the high polarization regime, by contrast, the dominant factor is the scarcity of minority spins that suppresses CAR only in the P configuration and hence makes $\delta R$  negative, as in the spin-independent interface model.

In Fig.~\ref{theory}(c) we see that our model reproduces the non-monotonic behavior of the experimental curves of the resistance in the two magnetization states. At low temperature $R^{\rm AP}$ is greater than $R^{\rm P}$ and, as the temperature increases, both curves dip down and then rise as Andreev processes become suppressed near $T_c$. The relative resistance [Fig.~\ref{theory}(d)] is greater than zero for the entire temperature range and, as in our experiment, becomes smaller as the temperature increases. While the qualitative features of our numerical result are in very good agreement with the experiment, we believe that in order to match the magnitude of $\delta R$ we need to employ more degrees of freedom in describing the scatterers, for example, by making $Z$ vary in two dimensions in future studies.

To fully test the applicability of our model, further experiments are required that vary the polarization of the ferromagnetic wires and the separation between them. On the theory front, incorporating induced spin-triplet correlations together with the spin-flip scatterers \citep{Almog11, Bergeret, Yokoyama07, Eschrig08} would provide a further ingredient for modeling additional features in SFS junctions. In the broader context, our studies reveal rich and unexpected physics, demonstrating that the manner in which pair correlations transfer to normal states is subtle and complex, and that a careful choice of interface parameters is required in constructing devices that hinge on spin physics in superconductivity-mediated nonlocal transport and electron entanglement.

This work is supported by the NSF Grants No. DMR06-05813 and No. DMR-0906521, by the University of Cincinnati (N.S.), and by the ARO Award No. W911NF-07-1-0464 (K.S.). N. S. and S. V. acknowledge the hospitality of the Aspen Center for Physics. We are grateful to M. Stehno for helpful discussions.

\end{document}